\documentstyle[aps,preprint]{revtex}
\begin{document}
\draft
\title{Tunneling effects on impurity spectral function in coupled asymmetric
quantum wires}
\author{Marcos R. S. Tavares}
\address{Faculdade de Tecnologia da Baixada Santista, CEETPS, Santos, SP 11045-908,\\
Brazil}
\author{G.-Q. Hai}
\address{Instituto de F\'{i}sica de S\~{a}o Carlos, Universidade de S\~{a}o Paulo,\\
S\~{a}o Carlos, SP 13560-970, Brazil}
\author{G. E. Marques}
\address{Departamento de F\'{i}sica, Universidade Federal de S\~{a}o Carlos, S\~{a}o\\
Carlos, SP 13565-905, Brazil}
\maketitle

\begin{abstract}
The impurity spectral function is studied in coupled double quantum wires at
finite temperatures. Simple anisotropy in the confinement direction of the
wires leads to finite non-diagonal elements of the impurity spectral
function matrix. These non-diagonal elements are responsible for tunneling
effects and result in pronounced extra peak in the impurity spectral
function up to temperatures as high as 20 K.
\end{abstract}


\section{Introduction}

From a theoretical point of view, it is well known that the Coulomb
interaction leads to the Luttinger liquid (LL) behavior of one-dimensional
(1D) electron systems. The LL is mostly characterized by the spin-charge
separation in the elementary excitation spectrum. \cite{Voit} However,
experimental results concerning inelastic light scattering due to both
charge-density (plasmon) and particle-hole (single-particle) excitations in
1D electron systems based on semiconductor quantum wires (SQWs) can be
interpreted within the random-phase approximation (RPA) in the framework of
the Fermi liquid (FL) theory. \cite{calleja} Among other issues, the Fermi
edge singularity \cite{tej} and the band gap renormalization \cite{byh1,8}
in these systems can also be successfully explained within the RPA. As a
matter of fact, the resonant inelastic light-scattering experiments in SQWs
have been analyzed using both the FL and LL theory. \cite{hu93,kra,wang2}
Recently, a new kind of tunneling spectroscopy in coupled parallel quantum
wires in GaAs-based semiconductors was able to check out the Luttinger
liquid behavior of the electrons in extremely clean systems at very low
temperatures. \cite{science} Differences between the Luttinger and Fermi
liquid behavior turned out to be noticeable in the one-electron spectrum.

So far, no definitive evidence for LL versus FL behavior has been obtained.
It is still a puzzle in clarifying the conditions under which the 1D
electron system embedded in SQW behaves as a LL. There have been theoretical
suggestions that mechanisms induced by impurity scattering, thermal
fluctuations as well as intersubband scattering can likely suppress the LL
behavior. \cite{hu93,mts,cap} Effects of the intersubband (or interwire)
excitations are believed to disable the LL behavior, provided the system
looses its strict one-dimensional character. Moreover, electron intra- and
intersubband relaxations through slightly impurity scattering channels could
restore the FL behavior. On the other hand, localized impurity (or defect)
induced effects in a 1D metallic systems in quantum wires have been
extensively studied \cite{jap} within both the FL and LL theories. These
impurity effects turn out to be related to critical phenomena in the
strongly correlated regime and to the X-ray absorption spectrum, where a
optical created valence hole is regarded as an impurity.

In this work we are interested in studying the impurity induced optical
properties in semiconductor quantum wires which can be helpful in
understanding the 1D properties of the electron systems. We focus on the
spectral function of a localized valence hole (or an spinless impurity) in a
two-component coupled double quantum wires at finite temperatures. The
considered interaction is of a localized state with a continuum described by
a coupled double 1D system with two subbands (or components) filled with
electrons. Two parallel wires in the $x$-direction, with zero thickness in
the $z$-direction, compose the system. In the $y$-direction, the two wires
are of 300 \AA\ and 200 \AA\ widths, respectively, separated by a 30-\AA\
AlGaAs barrier of 300 meV height. We assume zero thickness in the $z$%
-direction since the confinement in this direction is much stronger than
that in the $y$-direction. The energy gap between the first and the second
subbands is $\omega _{0}$ $\simeq $ 5.36 meV. We consider the impurity
localized at the center of the barrier. Such a system is also similar to a
type-II quantum wire heterostructures, where valence holes are localized
between two parallel 1D electron gases.

Our theoretical result indicates that the interwire tunneling induced
many-body effects in coupled asymmetric quantum wire systems lead to a
non-monotonic impurity spectral function. A pronounced extra peak appears in
the impurity spectral function up to 20 K.

The paper is organized as follows. In Sec. II we present the theory with the
working formulae; in Sec. III we provide our numerical results and
discussions; we conclude in Sec. IV with a summary.

\section{Theory}

\subsection{Interwire Coulomb potential}

The anisotropy in the direction perpendicular to the wires leads to a very
weak tunneling, but it is sensitive enough to produce important effects on
both the electron-electron (e-e) $V^{e-e}$ and the electron-impurity (e-i) $%
V^{e-i}$ Coulomb interactions. The asymmetry leads to nonzero off-diagonal
e-e interaction matrix elements of $V^{e-e}$. \cite{th01} These elements
represent e-e scattering in which only one of the electrons experiences
interwire (or intersubband) transition. In the symmetric situation (two
identical wires), however, these off-diagonal matrix elements vanish because
the wavefunctions of the ground and the first excited states are symmetrical
and anti-symmetrical, respectively, in the confinement direction.

On the other hand, the e-i interaction matrix element related to the
interwire transitions is defined as 
\begin{equation}
V_{12}^{e-i}(q)=\frac{2e^{2}}{\varepsilon _{0}}\int dy\phi
_{1}(y)K_{0}(q\left| y\right| )\phi _{2}(y),  \label{Vq}
\end{equation}
where $\varepsilon _{0}$ is the static dielectric constant, $e$ the electron
charge, $K_{0}(q\left| y\right| )$ the zeroth-order modified Bessel function
of the second order, and $\phi _{j}(y)$ the electron wavefunction in the
wire (or subband) $j$. The derivation of this potential is similar to that
in Ref. \cite{haiimp}, where its quasi-two-dimensional counterpart was
calculated. The Bessel function $K_{0}(q\left| y\right| )$ results from the
integral in the $x$-direction, which involves the external e-i Coulomb
potential 
\begin{equation}
V^{ext}(x,y)=\frac{e^{2}}{\varepsilon _{0}}\frac{1}{\sqrt{%
(x-x_{i})^{2}+(y-y_{i})^{2}}};  \label{vext}
\end{equation}
and the plane wave describing free electrons in that direction. It depends
obviously on the impurity position $x_{i}=y_{i}=0$, which has been chosen
here as the origin of the coordinates. The component $V_{12}^{e-i}(q)$
represents interwire (intersubband) transitions suffered by the electron as
it interacts with the localized impurity. This interaction induces
anisotropy (or tunneling) effects on the full impurity spectral function
that we investigate in this work. We mention that, for two identical wires, $%
V_{12}^{e-i}(q)=0$ due to the fact that the $\phi _{1}(y)$ and $\phi _{2}(y)$
are symmetrical and anti-symmetrical functions of $y$, respectively.

\subsection{Interwire impurity Greens function}

We model the system by an Anderson impurity localized below the continuum of
the two subbands ($j=1,2$) in the coupled quantum wires filled with
electrons. This approach has been used before in studying optical holes in
doped two-dimensional systems at zero temperature.\cite{hawho,hawho1,TTM} We
point out that the Hamiltonian presenting the interaction between the
impurity and the two-component 1D electron gas can be transformed into a
problem of a localized level with energy $\varepsilon _{i}$ and a
two-component non-interacting bosons (electron-hole pairs). Therefore, we
are dealing with an independent boson model which is exactly solvable. The
term in the Hamiltonian describing the boson coupling to the impurity is
written as 
\begin{equation}
H^{\prime }=d^{+}d\left[ \varepsilon _{i}+%
%TCIMACRO{\tsum}%
%BeginExpansion
\mathop{\textstyle\sum}%
%EndExpansion
_{\lambda q}V_{\lambda }^{e-i}(q)\left( b_{q\lambda }+b_{-q\lambda
}^{+}\right) \right] ,  \label{hamil}
\end{equation}
where $b_{\lambda q}^{+}$ ($b_{\lambda q}$) creates (destroys) a boson in
the state $\lambda \equiv (j,j%
%TCIMACRO{\UNICODE[m]{0xb4}}%
%BeginExpansion
{\acute{}}%
%EndExpansion
)$ with wavevector $q$. Here, $V_{\lambda }^{e-i}(q)$ is the
electron-impurity interaction matrix element. Notice that this coupling
occurs only when the impurity state is occupied and $d^{+}d=1.$

After a canonical transformation of the type $e^{S}H%
%TCIMACRO{\UNICODE[m]{0xb4}}%
%BeginExpansion
{\acute{}}%
%EndExpansion
e^{-S}$ with 
\begin{equation}
S=d^{+}d%
%TCIMACRO{\tsum}%
%BeginExpansion
\mathop{\textstyle\sum}%
%EndExpansion
_{\lambda q}\frac{V_{\lambda }^{e-i}(q)}{\omega _{\lambda }(q)}\left(
b_{-q\lambda }^{+}-b_{q\lambda }\right) ,  \label{Smatrix}
\end{equation}
the Hamiltonian (\ref{hamil}) can be rewritten as 
\begin{equation}
H^{\prime }=d^{+}d\left( \varepsilon _{i}-%
%TCIMACRO{\tsum}%
%BeginExpansion
\mathop{\textstyle\sum}%
%EndExpansion
_{\lambda }\Delta _{\lambda }\right) ,  \label{hamil2}
\end{equation}
where $\Delta _{\lambda }=%
%TCIMACRO{\tsum}%
%BeginExpansion
\mathop{\textstyle\sum}%
%EndExpansion
_{q}\left| \frac{V_{\lambda }^{e-i}(q)}{\omega _{\lambda }(q)}\right| ^{2}$
is the self-energy involving the frequency $\omega _{\lambda }(q)$ of the
boson in the state $\lambda $ with wavevector $q.$ One may then proceed to
obtain the impurity Greens function from the Hamiltonian (\ref{hamil2}).
After some algebra, an exact and closed form to the impurity Greens function
are obtained \cite{mahan} 
\begin{equation}
G(t)=-ie^{-it\left( \varepsilon _{i}-\Sigma _{\lambda }\Delta _{\lambda
}\right) }%
%TCIMACRO{\tprod}%
%BeginExpansion
\mathop{\textstyle\prod}%
%EndExpansion
_{\lambda }\exp \left[ -\Phi _{\lambda }(t)\right] ,  \label{green}
\end{equation}
where 
\begin{equation}
\Phi _{\lambda }(t)=\int_{0}^{\infty }\frac{d\nu }{\nu ^{2}}R_{\lambda }(\nu
)\left\{ \left[ n_{B}(\nu )+1\right] \left( 1-e^{-i\nu t}\right) +n_{B}(\nu
)\left( 1-e^{i\nu t}\right) \right\} ,  \label{phi(t)}
\end{equation}
with $n_{B}(\nu )$ being the Bose distribution function. The Feynman%
%TCIMACRO{\UNICODE{0xb4}}%
%BeginExpansion
\'{}%
%EndExpansion
s theorem on the disentangling of field operators is employed to determine $%
{\em G}(t)$. The benefit of treating the system within the independent boson
model comes when our working formulae are obtained without any approximation.

In Eq. (\ref{green}), $\Sigma _{\lambda }\Delta _{\lambda }$ is the
self-energy shift in relation to the impurity energy $\varepsilon _{i}.$
This shift is time independent and has no major effect on the total impurity
spectral function other than renormalizing the band-to-band threshold
transition energy $\omega _{T}=\varepsilon _{i}-\Sigma _{\lambda }\Delta
_{\lambda }$. For the sake of simplicity, we take $\omega _{T}=0$ throughout
this paper. We focus on the time-dependent self-energy term $\Phi _{\lambda
}(t)$ which requires the boson density of states \cite{mahan} 
\begin{equation}
R_{\lambda }(\nu )=\frac{1}{\pi }\int \frac{dq}{2\pi }\left| V_{\lambda
}^{s}(q,\nu )\right| ^{2}%
%TCIMACRO{\func{Im}}%
%BeginExpansion
\mathop{\rm Im}%
%EndExpansion
\left[ \Pi _{\lambda }(q,\nu )\right] ,  \label{R(w)}
\end{equation}
which determines the shape of the impurity spectral function. In Eq. (\ref
{R(w)}), $V_{\lambda }^{s}(q,\nu )$ is the screened e-i Coulomb potential,
and $\Pi _{\lambda }(q,\nu )$ is the non-interacting irreducible
polarizability function. The screened interaction $V_{\lambda }^{s}$ is
obtained from the generalized self-consistent equation, 
\begin{equation}
V_{\lambda }^{s}(q,\nu )=%
%TCIMACRO{\tsum}%
%BeginExpansion
\mathop{\textstyle\sum}%
%EndExpansion
_{\lambda 
%TCIMACRO{\UNICODE[m]{0xb4}}%
%BeginExpansion
{\acute{}}%
%EndExpansion
}\left[ \varepsilon _{\lambda ,\lambda 
%TCIMACRO{\UNICODE[m]{0xb4}}%
%BeginExpansion
{\acute{}}%
%EndExpansion
}(q,\nu )\right] ^{-1}V_{\lambda 
%TCIMACRO{\UNICODE[m]{0xb4}}%
%BeginExpansion
{\acute{}}%
%EndExpansion
}^{e-i}(q),  \label{Vscreened}
\end{equation}
with the dielectric matrix 
\begin{equation}
\varepsilon _{\lambda ,\lambda 
%TCIMACRO{\UNICODE[m]{0xb4}}%
%BeginExpansion
{\acute{}}%
%EndExpansion
}=\delta _{\lambda \lambda 
%TCIMACRO{\UNICODE[m]{0xb4}}%
%BeginExpansion
{\acute{}}%
%EndExpansion
}-V_{\lambda \lambda 
%TCIMACRO{\UNICODE[m]{0xb4}}%
%BeginExpansion
{\acute{}}%
%EndExpansion
}^{e-e}\Pi _{\lambda 
%TCIMACRO{\UNICODE[m]{0xb4}}%
%BeginExpansion
{\acute{}}%
%EndExpansion
}  \label{diel}
\end{equation}
being written within the RPA.

We point out that the formula for the density of states $R_{\lambda }(\nu )$
has been conveniently chosen in order to take into account elementary
excitations (or screening) occurring in the experiment. The quantity $%
R_{\lambda }(\nu )$ is also playing a role of a density of states for the
electron-electron interaction provided it is determined by collective
excitations (entering in the calculation via the screened e-i potential $%
V_{\lambda }^{s}(q,\nu )$), which remain in the particle-hole excitation
continuum where $%
%TCIMACRO{\func{Im}}%
%BeginExpansion
\mathop{\rm Im}%
%EndExpansion
\left[ \Pi _{\lambda }(q,\nu )\right] \neq 0.$ It will then depend on the
electron density $n_{e}$ in the sample, since the polarizability $\Pi
_{\lambda }$ does. Therefore, in calculating $R_{\lambda }(\nu )$, we are
dealing with a boson density of states, {\em which have been mapped onto
elementary excitations in the real electron system. }

The elementary excitation spectra of a Fermi system embedded in the biwire
structure present the well-known optical, $\omega _{+}(q),$ and acoustic, $%
\omega _{-}(q)$, plasmon modes corresponding to the in-phase and
out-of-phase charge density fluctuations, respectively. \cite{9} The
dispersion relations of these modes are obtained through the equation $\det
\left| \varepsilon _{\lambda \lambda 
%TCIMACRO{\UNICODE[m]{0xb4}}%
%BeginExpansion
{\acute{}}%
%EndExpansion
}(q,\omega )\right| =0$.

We devoted special attention in this work to the element $R_{12}(\nu ),$
which is interpreted as the rate of creating interwire electron-hole pairs
(bosons) with electron (hole) being in the wider (narrower) wire. The
tunneling between the wires is responsible for $V_{12}^{e-i}(q)$ and, as a
consequence, for $R_{12}(\nu )$. It then leads to the interwire impurity
spectral function defined as 
\begin{equation}
A_{12}(\omega )=-2%
%TCIMACRO{\func{Im}}%
%BeginExpansion
\mathop{\rm Im}%
%EndExpansion
\left[ 
%TCIMACRO{\tint }%
%BeginExpansion
\textstyle\int%
%EndExpansion
dte^{i\omega t}g_{12}(t)\right] ,  \label{Aw}
\end{equation}
where 
\[
g_{12}(t)=-i\exp \left[ -\Phi _{12}(t)\right] 
\]
is the finite temperature impurity Greens function involving the electronic
transitions between the wires. The impurity spectral function in 3D systems
at zero temperature diverges as power-law at $\omega \rightarrow \omega _{T}$%
. This divergency is due to the so-called orthogonality catastrophe
involving the overlap between the initial (without impurity) and final (with
impurity) many-body states. \cite{14} Here, the Bose distribution function $%
n_{B}(\omega )$, governing the particle-hole excitations at finite
temperatures, plays an important role being responsible for the behavior of $%
A_{12}(\omega )$ as $\omega \rightarrow 0$.

\section{Numerical results}

In Fig. 1 we show the $R_{12}(\nu )$ at different temperatures for the total
charge density $n_{e}=10^{6}$ cm$^{-1}$. As discussed above, Eq. (\ref{R(w)}%
) indicates that the screened e-i potential $V_{12}^{s}(q,\nu )$ in the
interwire particle-hole continuum, where $%
%TCIMACRO{\func{Im}}%
%BeginExpansion
\mathop{\rm Im}%
%EndExpansion
\left[ \Pi _{12}(q,\nu )\right] \neq 0,$ determines the interwire density of
states $R_{12}(\nu )$. On the other hand, the weak interwire particle-hole
excitations in the present system only result in partially Landau damping on
the plasmon modes as we discussed in Ref. \cite{th01}. Therefore, the
survived optical $\omega _{+}(q)$ and acoustic $\omega _{-}(q)$ plasmon
modes in this region are of dominant contribution to $V_{12}^{s}(q,\nu )$
and, consequently, in determining the $R_{12}(\nu )$. In order to understand
that, we show in Fig. 2 the intra- and interwire particle-hole excitation
spectra along with the plasmon dispersion relations of the optical and the
acoustic modes in the system with $n_{e}=10^{6}$ cm$^{-1}$ at zero{\bf \ }%
temperature. The corresponding\ $R_{12}(\nu )$ is given by the solid curve
in Fig. 1. Clearly, there are well defined borders of the interwire
particle-hole continuum at zero temperature. The well shaped $R_{12}(\nu )$
follows the optical and acoustic branches entering and leaving each
continuum region. The acoustic (optical) mode enters and leaves the lower
interwire particle-hole continuum region at $\omega _{-}$ $\simeq 3.0$ and $%
3.8$ meV ($\omega _{+}$ $\simeq 3.7$ and $4.3$ meV), respectively. The
optical one enter the higher interwire particle-hole continuum region at $%
\omega _{+}$ $\simeq 6.3$ meV. In comparison with the plasmon dispersions,
our calculation indicates that the $R_{12}(\nu )$ is dominated by the
acoustic and the optical plasmon branches which remain in the interwire
particle-hole continuum. We address that, in calculating $R_{12}(\nu )$, we
are dealing with an interwire boson density of states, which have been
mapped onto elementary excitations in the Fermi system.

For the sake of completeness, we show the intrawire element $R_{11}(\nu )$
in the inset of Fig. 1 for the same temperature range. The rate $R_{11}(\nu
) $ is dominated by the intrawire particle-hole excitations. But the
partially damped acoustic plasmon mode in the intrawire particle-hole
excitation continuum has a small contribution to $R_{11}(\nu )$ in our
asymmetric system. We also show that, quantitatively, the $R_{12}(\nu )$ is
larger than the $R_{11}(\nu )$\ at very low temperatures. Consequently, it
dominates the full impurity spectral function at low temperatures. In
increasing temperature, $R_{12}(\nu )$\ ($R_{11}(\nu )$) decreases
(increases) and the interwire effects carried by $R_{12}(\nu )$ are
suppressed by $R_{11}(\nu )$. Moreover, the function $R_{12}(\nu )$ is of
different behavior in comparing with the $R_{11}(\nu )$ which is roughly a
slowly varying function. The abrupt increase in $R_{12}(\nu )$ leads to a
specific structure in the impurity spectral function. We also found that the
elements $R_{22}(\nu )$ and $R_{21}(\nu )$ are of negligible contribution
due to the small probability of creating electron-hole pairs whose electrons
are found in the narrower wire.

In Fig. 3, the dotted and dashed lines show $A_{jj%
%TCIMACRO{\UNICODE[m]{0xb4}}%
%BeginExpansion
{\acute{}}%
%EndExpansion
}(\omega )$ for $(j,j%
%TCIMACRO{\UNICODE[m]{0xb4}}%
%BeginExpansion
{\acute{}}%
%EndExpansion
)=(1,1)$ and $(j,j%
%TCIMACRO{\UNICODE[m]{0xb4}}%
%BeginExpansion
{\acute{}}%
%EndExpansion
)=(1,2)$, respectively, at $T=2$ K. The full spectrum (solid line) should be
observed experimentally via photoemission or photocurrent spectroscopy and
are proportional to the impurity density of states involving elementary
excitations mostly due to the wider quantum wire. The interwire tunneling
induced effects as discussed above reflect on the non-diagonal element $%
A_{12}(\omega )$. We see structures in the spectral function $A_{12}(\omega
) $ in which dip and peak correspond to the behavior of $R_{12}(\nu )$.
These structures represent the contribution coming from interwire boson
excitations, which decrease as the temperature increases.

In Fig. 4(a), we show $A_{12}(\nu )$ at temperature $T=2$ K for different
charge densities $n_{e}$. The figure clearly shows the effects due to
interwire particle-hole excitations and their dependence on the total charge
density in the sample. As $n_{e}$ decreases, the narrower wire becomes less
populated and the shoulders in the figures get affected.\ In Fig. 4(b), we
show the temperature evolution of $A_{12}(\omega )$ for $n_{e}=10^{6}$ cm$%
^{-1}$. Although the temperature suppresses the interwire boson excitations,
the effects due to interwire interaction is observable in our calculation up
to $T=30$ K. Moreover, the Boson distribution function $n_{B}(\omega ),$
governing the interwire self-energy propagator $\Phi _{12}(t)$ at finite
temperatures, leads to the divergency in $A_{12}(\omega )$ for $\omega
\rightarrow 0$.

\section{Summary}

In summary, we have studied the non-diagonal (interwire) element of the
spectral function matrix of a localized impurity in a coupled double quantum
wires. We show that the tunneling between the two asymmetric wires induces
interwire particle-hole excitation effects leading to a pronounced peak in
the impurity spectral function up to 20 K which should be observable in the
photoemission or photocurrent spectroscopy. We finally remark that
experimental observations of these effects could serve as a signature of
one-dimensional FL in the proposed system.

\section{Acknowledgments}

This work is supported by {\em FAPESP}, the research foundation agency of
the state of S\~{a}o Paulo, Brazil.

\begin{figure}[tbp]
\caption{The rate $R_{12}(\protect\nu )$ for $n_{e}=10^{6}$ cm$^{-1}$ at
different temperatures: $T=0$ K (solid line), $T=5$ K (dotted line), $T=10$
K (short-dashed line), $T=20$ K (dashed line), and $T=25$ K (dash-dotted
line). The inset shows the intrawire rate $R_{11}(\protect\nu )$ for the
same temperature values. }
\label{fig1}
\end{figure}

\begin{figure}[tbp]
\caption{ Dispersion relations of the optical $\protect\omega _{+}(q)$ and
acoustic $\protect\omega _{-}(q)$ plasmon modes with $n_{e}=10^{6}$ cm$^{-1}$%
. Shaded areas indicate the intra- and inter-wire particle-hole excitation
continua. Inset shows the wire structure and the wavefunctions of the ground
and first excited state. }
\label{fig2}
\end{figure}

\begin{figure}[tbp]
\caption{ The elements $A_{jj{\acute{}}}(\protect\omega )$ at temperature
T=2 K. Here, $n_{e}=10^{6}$ cm$^{-1}$. }
\label{fig3}
\end{figure}

\begin{figure}[tbp]
\caption{ (a) The interwire element $A_{12}(\protect\omega )$ at temperature
T=2 K for different electron densities $n_{e}$. (b) The interwire element $%
A_{12}(\protect\omega )$ for $n_{e}=10^{6}$ cm$^{-1}$ at different
temperatures. }
\label{fig4}
\end{figure}

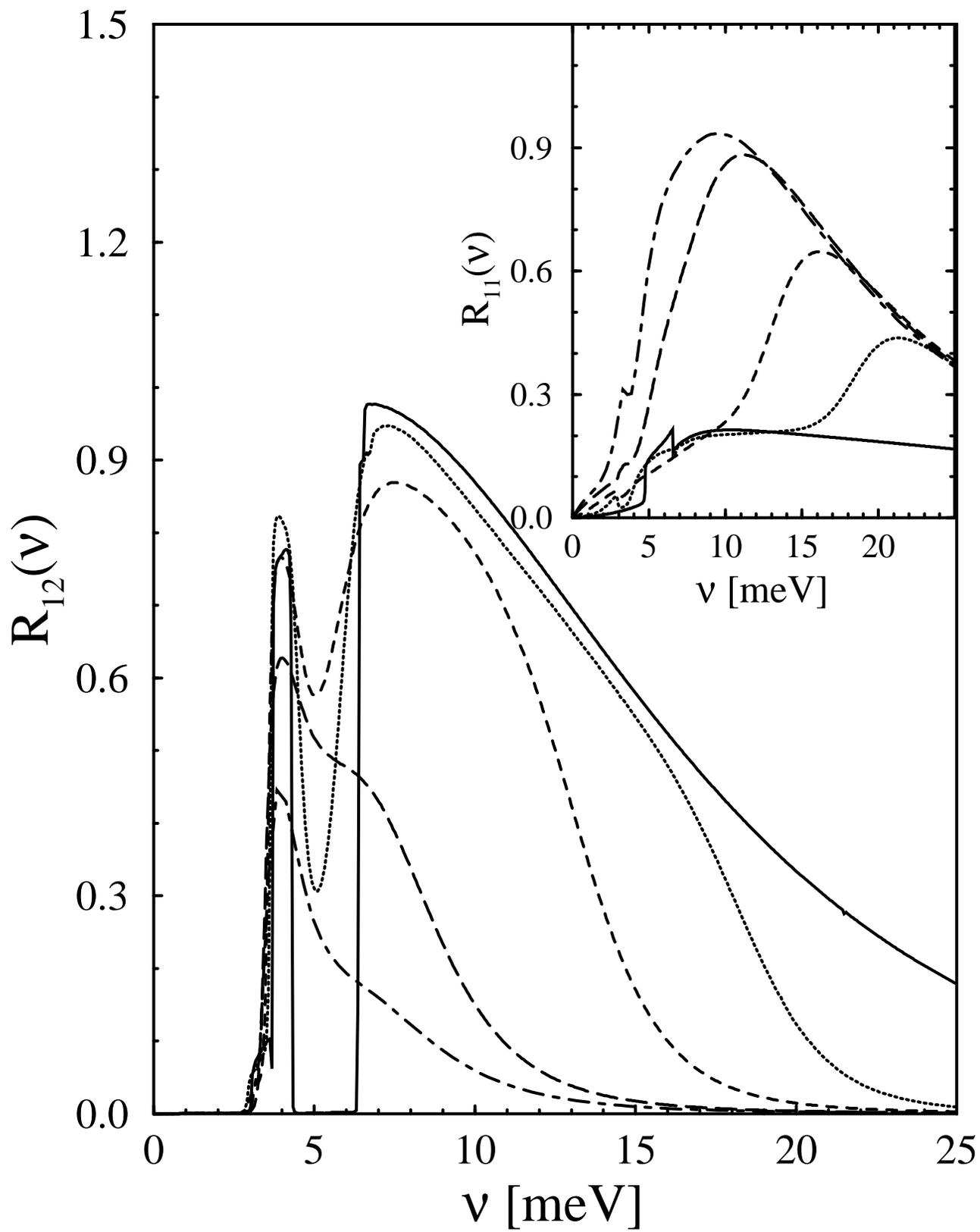
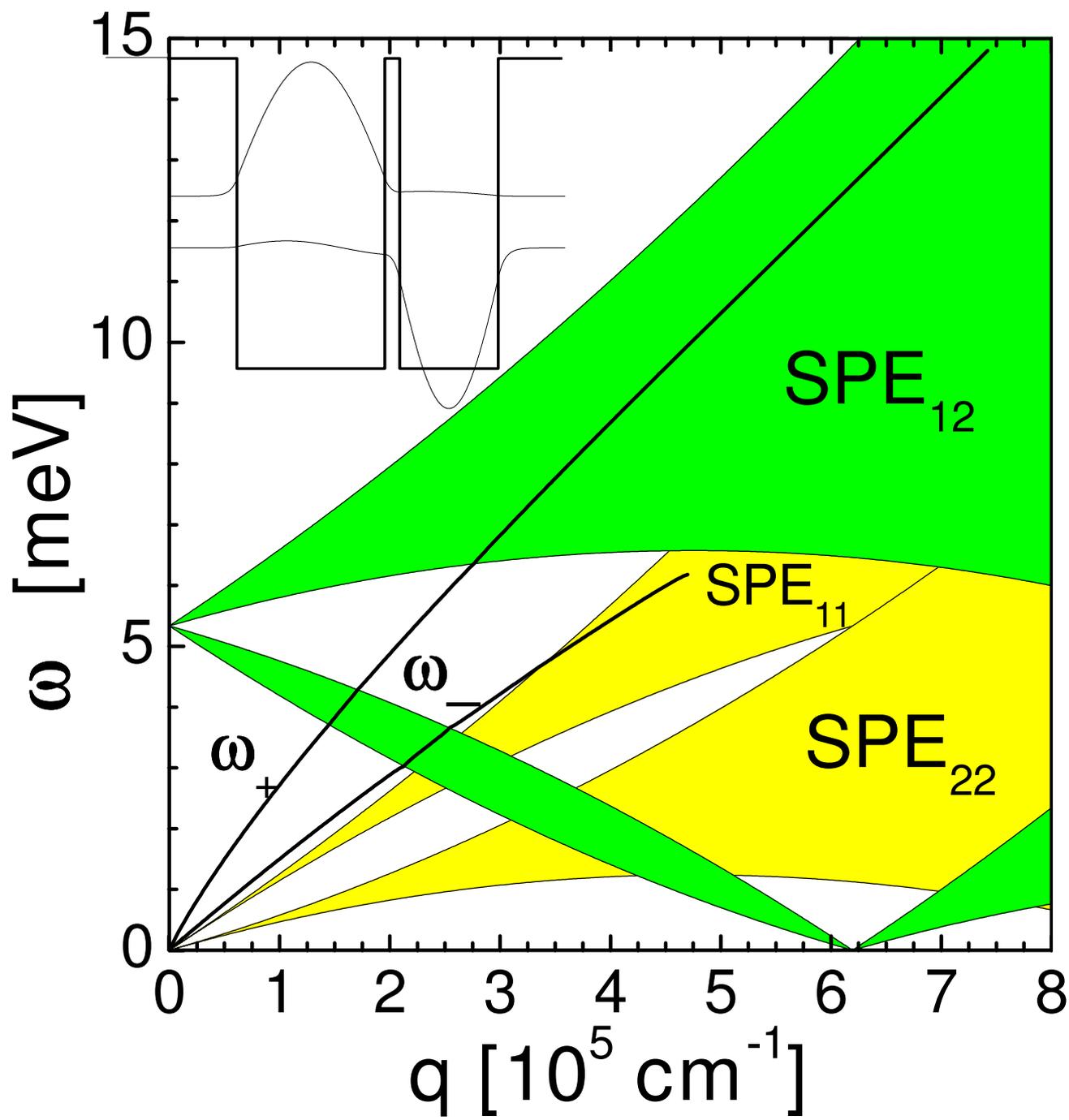
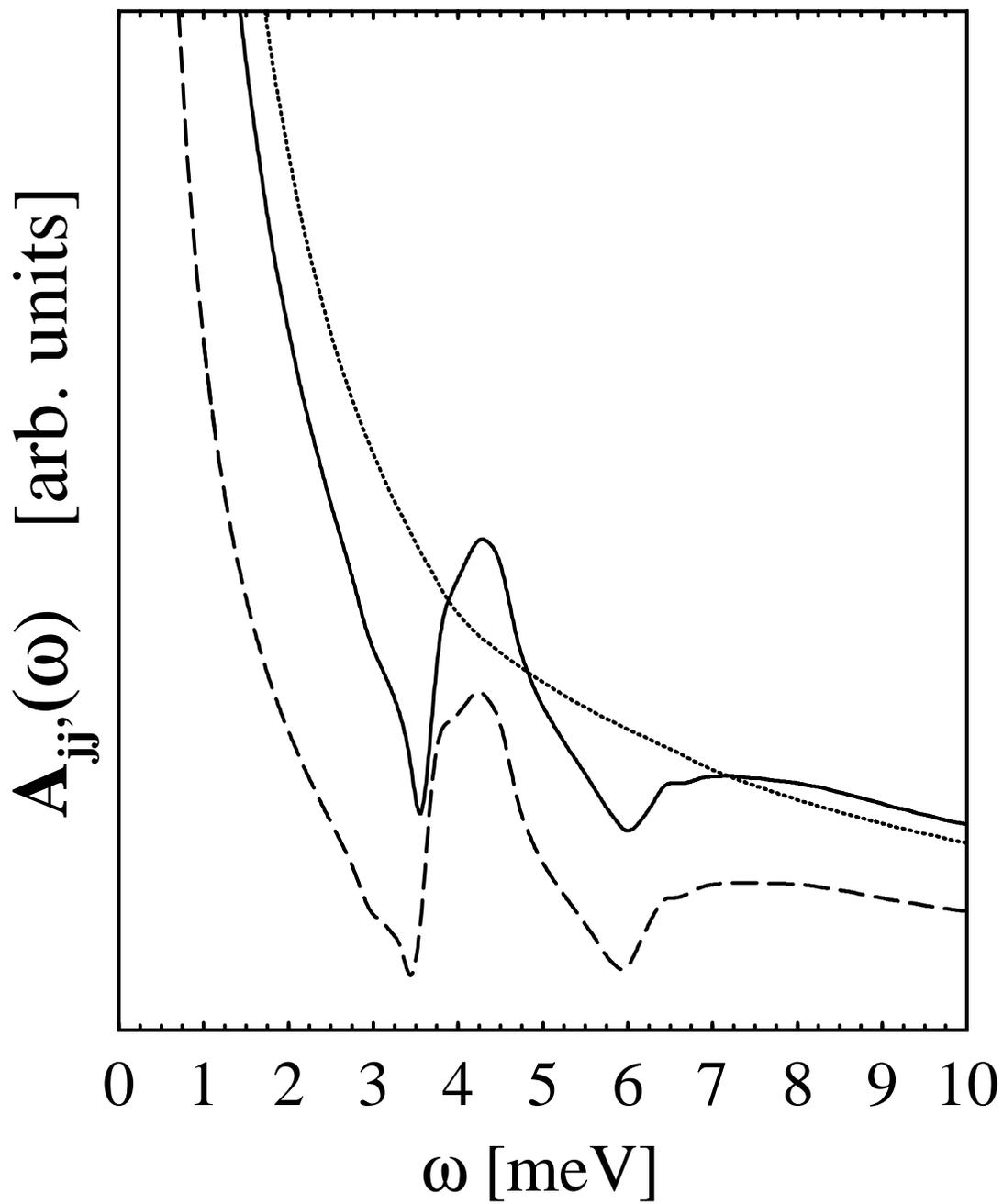
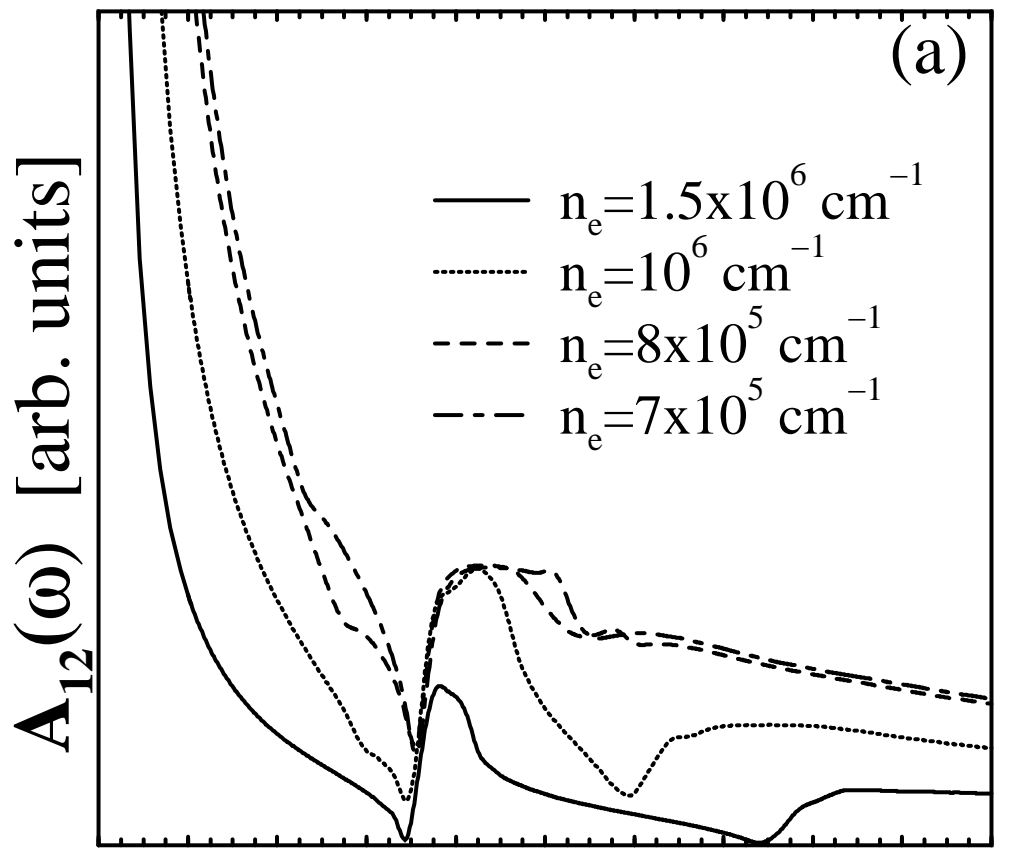
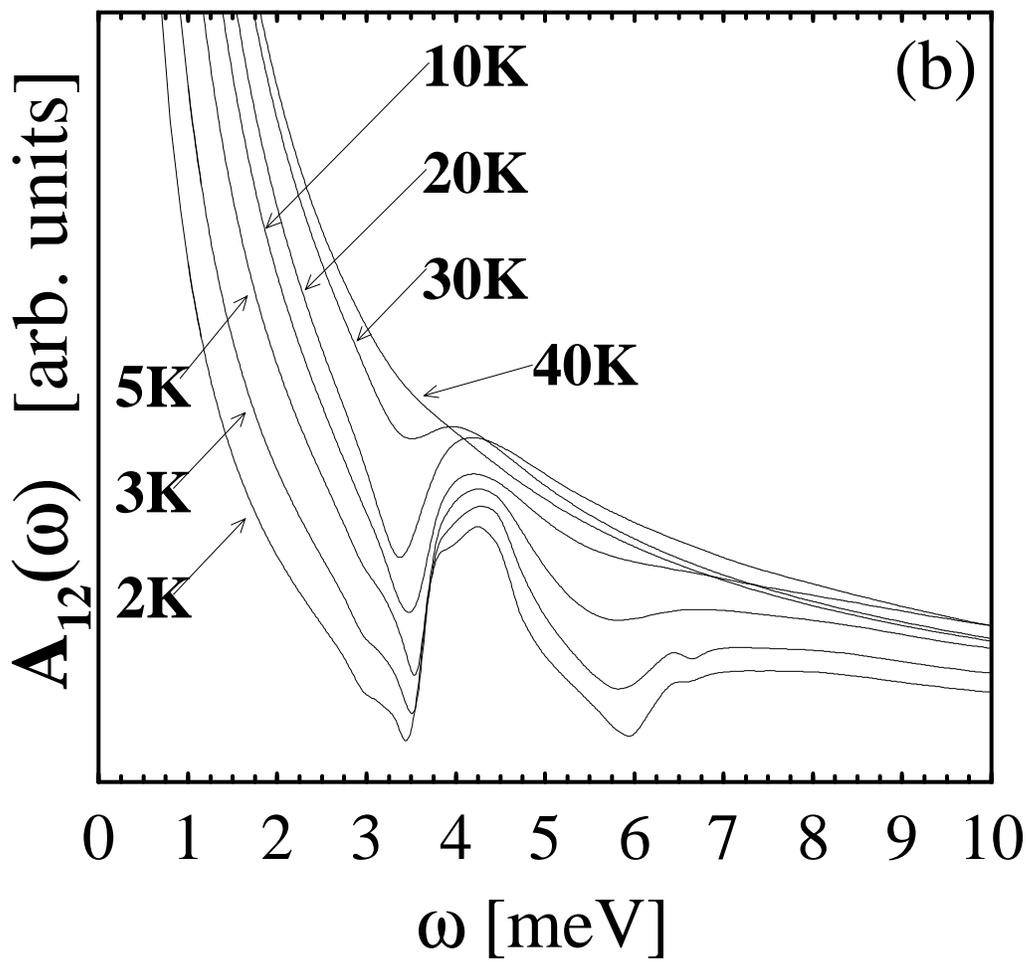
\end{document}